\documentclass[aps,prl,twocolumn,groupedaddress]{revtex4}
\usepackage{graphicx}
\usepackage{amsmath}

\begin{document}


\title{Evaluation of the Permutational Structure of Quantum Gases at Finite Temperature}


\author{Ryan Springall}
 \email[Corresponding author ]{ryan.springall@rmit.edu.au}
 \affiliation{Department of Physics, RMIT University Australia}
\author{Manolo Per}
 \affiliation{Department of Physics, RMIT University Australia}
\author{Ian Snook}
 \affiliation{Department of Physics, RMIT University Australia}


\date{\today}

\begin{abstract}
Proposed is an alternative method for permutational sampling in
quantum gases using the path integral formulation of statistical
mechanics. It is shown that in principle we are able to use two
operators which enable us to construct a Markov chain through a
graph of the irreducible representation of the symmetric group. As
an illustration of this method, a test calculation of four particles
in a harmonic trap is performed.
\end{abstract}


\maketitle


\section{Introduction}

Irrespective of their experimental realisation, the theoretical
existence of condensed phases of noninteracting media is indeed
remarkable, and the mathematical structures with which they are
imbued can lead to a variety of interesting observations. It is not
difficult to show that a system of noninteracting indistinguishable
particles with bosonic symmetry will undergo a second order phase
transition if the spatial dimension is greater than two. For
$d\leq{2}$ it has been rigorously been shown that a bosonic system
will not undergo phase transition for $T>0$
\cite{Hohenberg-1967,BEC-book}. Theoretically however it was found
that this is dependent on the form of the confining
potential\cite{BEC-2D-Theory} confirmed by evidence of Bose-Einstein
condensates (BEC's) having recently been observed in quasi-2d
traps\cite{BEC-2D}.

The experimental observation of BEC's in systems with both positive
and negative s-wave scattering
length\cite{Exp_BEC,BEC-Lithium,BEC-Sodium}, and the rapid expansion
of novel trapping potentials in BEC's such as optical and magnetic
lattices \cite{Optical-Lattice,Optical-Lattice2,Magnetic-Lattice}
and quasi-one and two dimensional harmonic traps
\cite{Cigar-trap,Cigar-trap2,BEC-2D}, has given a renewed impetus to
the theoretical description of low temperature atomic gases.

Of the theoretical approaches, path integral Monte Carlo(PIMC)
techniques stand as one of the most useful non-perturbative methods,
and is the only method able to produce exact properties of systems
at finite temperature, via sampling of the thermal density matrix.
One of the ubiquitous problems in all approaches to evaluating the
partition function for bosonic systems is a transparent way of
sampling the permutation space inherent in all problems involving
indistinguishable particles. Previous approaches such as those in
\cite{Ceperley-1995,Boninsegni-2005,Holzmann-1999} advocate either
the sampling of some cyclic subset of $S_{n}$, the symmetric group
of order $n!$, typically up to $C_{4}$ or $C_{5}$, or approximate
the interacting partition function structure to that of a
noninteracting system, which is only strictly valid for systems
which are weakly coupled.

In this paper we wish to make some remarks regarding the
permutational structure of the partition function for bosonic
particles. We present an algorithm which allows us to construct a
Markov chain through permutation space for an explicit probability
measure which maps between conjugacy classes of the symmetric group.
It is partially inspired by the recent work regarding Markov
processes on Young tableaux
\cite{Borodin-2000,Borodin-2004,Borodin-Markov,Okounkov-2000} and
the connection between irreducible representations of the symmetric
group and the analytic form of the bosonic partition function for
noninteracting systems. Unlike a previous method
\cite{Holzmann-1999}, which is restricted to weakly interacting
systems, we propose that this method samples the ring configuration
structure at any temperature without any \textit{a priori} knowledge
of the noninteracting counterpart. This enables us to quantify ring
configuration probabilities over a larger temperature scale and
evaluate the relative importance of certain structures to the
partition function.

The aims of this paper are $(1)$ to give a rigorous formulation of
the role of permutation cycles in the canonical ensemble partition
function via a correspondence with irreducible representations of
the symmetric group, and $(2)$ to propose a practical scheme for
computational implementation within PIMC. As an illustration we
perform calculations on noninteracting bosons in a 1-dimensional
harmonic potential, the results of which are presented in sec. IV, a
system which has a very well understood theoretical basis.

This paper is organised as follows. In Sec. II we provide the
theoretical outline for the path integral formulation of statistical
mechanics in the canonical ensemble and derive a form of the
partition function based upon the conjugacy classes of the symmetric
group of $n$ elements. In Sec. III we outline the basis for
effective evaluation of the discrete permutational space by
considering random growth of partitions of the symmetric group using
one loop operators. We are able to stochastically sample the graph
of partitions and show that this theoretically will enable us to
determine loop configuration probabilities as a function of
temperature.

\section{Path Integral Statistical Mechanics}
The basis of the path integral formulation of statistical mechanics
is the decomposition property of the many body density matrix
$\rho\left(R,R';\beta\right)$, and the resulting classical
configuration integral for calculating the properties of quantum
systems \cite{Feyn-Stat-Mech,Ceperley-1995}

\begin{equation}
\label{density.decomp}
\rho(R,R';\beta)=\int{dR_{1}\rho(R,R_{1};\beta/2)\rho(R_{1},R';\beta/2)}
\end{equation}

where

\begin{eqnarray}
\label{density.mat}
\rho(R,R';\beta)&=&\langle{R|e^{-\beta\mathcal{H}}|R'}\rangle
\end{eqnarray}

and $|R\rangle=|{x^{1},x^{2},...,x^{N}}\rangle$ denotes the
coordinate representation of the full $N$ particle Hilbert space.
The Hamiltonian operator $\mathcal{H}$ in general contains q-number
quantities representing particle kinetic and potential terms and Eq.
(2) represents that of an interacting many body quantum problem for
distinguishable (Boltzmannian) particles. The partition function is
given as the trace of the density matrix

\begin{eqnarray}
\nonumber Z&=&Tr\left(\rho\right)\\
\nonumber &=&\int{dR\langle{R|e^{-\beta\mathcal{H}}|R}\rangle}\\
&=&Tr\left(e^{-\beta\mathcal{H}}\right),
\end{eqnarray}

the final line indicating that it is independent of the
representation. The convolution property of the density matrix is
exact and gives us a representation of the density matrix at
temperature T as the convolution of two density matrices at
temperature 2T, effectively mapping low temperature problems to that
of higher temperatures. Using the Trotter formula,
$exp\left(-\beta\mathcal{H}\right)=exp\left(-\tau\mathcal{H}\right)^{M}$,
where $\tau=\beta/M$, we may perform this operation an arbitrary
number of times to obtain the expression

\begin{eqnarray}
Z&=&\prod^{M}_{i=0}\left[\int{dR_{i}}{\rho(R_{i+1},R_{i};\beta/M)}\right],
\end{eqnarray}

with the trace condition requiring that $R_{M+1}=R_{0}$. In the
limit that M tends to infinity, $\tau\rightarrow{0}$ and we obtain
an expression for the partition function which is functionally
dependent on the classical action. The path integral expression is

\begin{eqnarray}
\label{eq.boltz_part}
Z&=&\oint^{R(\beta)=R(0)}_{R(0)}{\mathcal{D}R{e^{-\beta{S}}}}
\end{eqnarray}

where

\begin{eqnarray}
\oint{\mathcal{D}R}&=&\lim_{M\rightarrow\infty}\prod^{M}_{i=0}\left[\int{dR_{i}}\right]
\end{eqnarray}

is the functional integration measure. The dynamical quantity of
interest, the Euclidean action, is now rendered a c-number quantity
on which the partition functionally is dependent.

For the discrete form of the partition function given in Eq.(4), the
procedure of taking the trace of a convolution of M density matrices
is to represent particles as a set of M classically interacting
beads interacting via a temperature dependent harmonic force,
forming a closed 'necklace'. This is sometimes referred to as the
polymer isomorphism \cite{Ceperley-1995}. It should be noted however
that the inter-particle potential only acts between beads of
corresponding time-slices $\left(M\right)$
\cite{Ceperley-1995,Feyn-Stat-Mech}. As the number of time slices
tends to infinity particles are represented as continuous closed
loops. For an extensive review of path integrals in this context,
the reader is referred to \cite{Ceperley-1995}.

The formal evaluation of Eq.\ref{eq.boltz_part} for noninteracting
particles, gives the partition function to be the product of the
single particle partition function, i.e.

\begin{equation}
Z_{N}[\beta] = Z_{1}[\beta]^{N}.
\end{equation}

When considering ensembles of indistinguishable particles one is
required to permutate over particle labels. In the discretised form
of the partition function we need only consider permutations over
particles in the final time slice \cite{Ceperley-1995}, i.e.

\begin{eqnarray}
\nonumber \rho_{B}\left(R,R';\beta\right)&=&\frac{1}{N!}\sum_{\phi\in{S_{N}}}\rho\left(R,\phi{R'};\beta\right)\\
\nonumber Z_{B}&=&\frac{1}{N!}\sum_{\phi\in{S_{N}}}\prod^{M-1}_{i=0}\left[\int{dR_{i}}{\rho(R_{i+1},R_{i};\beta/M)}\right]\times\\
& & \int{dR_{M}\rho\left(R_{0},\phi{R_{M}},\beta/M\right)}.
\end{eqnarray}

For noninteracting systems of identical particles, the convolution
property of the density matrix leads to a decomposition and the
partition function becomes the product of density matrices at
various temperatures. Consider a system of two particles with the
Hamiltonian operator

\begin{eqnarray}
\mathcal{H}&=&\mathcal{H}_{1}+\mathcal{H}_{2}
\end{eqnarray}

such that $\left[\mathcal{H}_{1},\mathcal{H}_{2}\right]=0$. In this
case the Boltzmannian 2-body density matrix will be
$\rho(x_{1}x_{2},x'_{1}x'_{2};\beta)=\rho(x_{1},x'_{1};\beta)\rho(x_{2},x'_{2};\beta)$.
The partition function for the bosonic case will be

\begin{eqnarray}
\nonumber Z_{B}[\beta]&=&\frac{1}{2}\int{dx_{1}dx_{2}}\rho(x_{1},x_{1};\beta)\rho(x_{2},x_{2};\beta)\\
&&+\frac{1}{2}\int{dx_{1}dx_{2}}\rho(x_{1},x_{2};\beta)\rho(x_{2},x_{1};\beta)
\end{eqnarray}

Using Eq.(1) for the second integral in Eq.(10), the partition function
can be written exactly as

\begin{eqnarray}
Z_{B}[\beta]&=&\frac{1}{2}\left[Z_{1}[\beta]^{2}+Z_{1}[2\beta]\right],
\end{eqnarray} $Z_{1}$ denoting the single particle partition function. As is
frequently noted, the partition function of statistical mechanics is
related to the propagator of quantum theory via rotation of time to
the imaginary axis \cite{Kleinhert-Path-Int}, and so the effect of
permuting particle labels in the two particle case is to wind the
trajectory twice around imaginary time

A general expression can be obtained for the N-body noninteracting
partition function, involving the irreducible representations of the
symmetric group. Consider the symmetric group, denoted by $S_{n}$,
which is the group of all permutations on n objects, with the order
being $|S_{n}|=n!$. The number of conjugacy classes of $S_{n}$ is
equal to the number-theoretic partition function
$P(n)$\cite{Discrete-Math}, with the property

\begin{equation}
\bigcup^{P(n)}_{i=1}\phi^{i}=S_{n}.
\end{equation}

Here $\phi^{i}$ is the i$^{th}$ conjugacy class. The conjugacy
classes can be graphically represented via the standard Young
tableaux, however for this case it is instructive to consider
diagrams as shown in Fig.(1). This originates from the
representation of particles in path integral statistical mechanics
as kinetic strings. The action of elements of $S_{n}$ on these
strings is to attach each string starting position ($\tau=0$) with
all possible final positions and integrate, giving the partition
function. It should be noted that this is exactly true for the
interacting as well as the noninteracting case.

\par Consider $S_{4}$ which contains a total of 5 partitions and 24
elements, which is shown in Fig 1. From this is can be seen that
elements of a particular conjugacy class are topologically
equivalent. We can identify Young tableaux and partitions of n with
the weakly decreasing sequence

\begin{figure}
  \includegraphics[width=5cm]{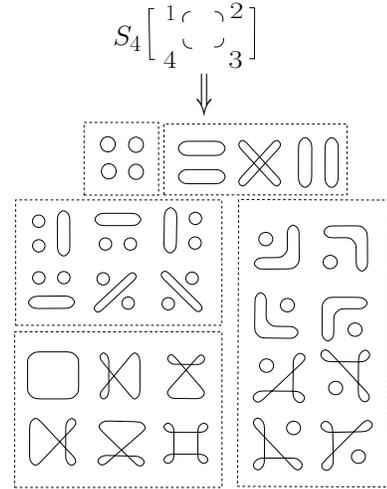}\\
  \caption{Representation of the Conjugacy Classes of $S_{4}$}\label{fig.loops}
\end{figure}

\begin{eqnarray}
\nonumber\bar{\phi}^{i} & = & \{ \bar{\lambda}^{i}_{1},\bar{\lambda}^{i}_{2},...,\bar{\lambda}^{i}_{\rho}\},\\
\end{eqnarray} each element having the property that
$\bar{\lambda}^{i}_{1}\geq\bar{\lambda}^{i}_{2}\geq...\geq\bar{\lambda}^{i}_{\rho}$,
or as the strongly decreasing sequence

\begin{eqnarray}
\phi^{i} & = & \{\left(\lambda^{i}_{1}\right)^{k_{i,1}},\left(\lambda^{i}_{2}\right)^{k_{i,2}},...,\left(\lambda^{i}_{\gamma}\right)^{k_{i,\gamma}}\}
\end{eqnarray}

where $\lambda^{i}_{1}>\lambda^{i}_{2}>...>\lambda^{i}_{\gamma}$. Here
$k_{i,j}$ will be denoted as the multiplicity of a particular loop
size. The dimension of each conjugacy class is equal to the number
of Young tableaux with shape $\phi^{i}$ which is given by

\begin{equation}
\textrm{dim}\left(\phi^{i}\right)=\frac{n!}{\prod_{j=1}^{\gamma}\left(\left(\lambda_{j}^{i}\right)^{k_{i,j}}k_{i,j}!\right)}
\end{equation}

For the purpose of this paper we need to identify two numbers
$|\bar{\phi}^{i}|=\rho$, the number of elements in a partition,
which is equal to the number of rows on the equivalent Young
tableaux and $|\phi^{i}|=\gamma$, the number of distinct elements.
For example if we denote $\Phi$ as the set of all Young tableaux
with n boxes, for the case of $S_4$ we have the conjugacy classes
represented by the partitions

\begin{eqnarray}
\nonumber\Phi&=&\{ \{1,1,1,1\},\{2,1,1\},\{2,2\},\{3,1\},\{4\}\}\\
\nonumber&=&\{\{1^{4}\},\{2,1^{2}\},\{2^{2}\},\{3,1\},\{4\}\}\\
&=&\{\phi^{1},\phi^{2},\phi^{3},\phi^{4},\phi^{5}\},
\end{eqnarray}

with the dimension of each conjugacy class given as

\begin{eqnarray}
\nonumber{\textrm{dim}}\left(\Phi\right)&=&\{\frac{24}{1^4.4!},\frac{24}{2.1^{2}.2!},\frac{24}{2^{2}.2},\frac{24}{3.1},\frac{24}{4}\}\\
&=& \{1,6,3,8,6\}.
\end{eqnarray}

With this notation, we are able to write down a general expression
for the partition function of a noninteracting bosonic system of n
particles, which is

\begin{equation}
\label{eq.partition}
Z_{B}[\beta]=\frac{1}{N!}\sum_{i=1}^{P(n)}\left(\textrm{dim}\left(\phi^{i}\right)\prod_{j=1}^{|\phi^{i}|}Z_{1}[\lambda^{i}_{j}\beta]^{k_{i,j}}\right)
\end{equation}

where the superscript $k_{i,j}$ denotes raising the power of the
partition function to the multiplicity of this particular loop with
cycle size $\lambda^{i}_{j}$. To obtain an understanding of Eq.(18),
let us again consider the example of $S_{4}$ for which $P(4)=5$.
Written out fully, the partition function takes the form

\begin{eqnarray}
\label{4.bosonic} \nonumber 24\times Z_{B} & = &
\prod_{j=1}^{|\phi^{1}|}Z_{1}\left[\lambda^{1}_{j}\beta\right]^{k_{1,j}}+
6\prod_{j=1}^{|\phi^{2}|}{Z_{1}}\left[\lambda^{2}_{j}\beta\right]^{k_{2,j}}+\\
\nonumber  &   &
3\prod_{j=1}^{|\phi^{3}|}Z_{1}\left[\lambda^{3}_{j}\beta\right]^{k_{3,j}}+
8\prod_{j=1}^{|\phi^{4}|}Z_{1}\left[\lambda^{4}_{j}\beta\right]^{k_{4,j}}+\\
\nonumber  &   &
6\prod_{j=1}^{|\phi^{5}|}Z_{1}\left[\lambda^{5}_{j}\beta\right]^{k_{5,j}}\\
\nonumber & = &
 Z_{1}[\beta]^{4}+6Z_{1}[2\beta]Z_{1}[\beta]^{2}+3Z_{1}[2\beta]^{2}\\
&  & +8Z_{1}[3\beta]Z_{1}[\beta]+6Z_{1}[4\beta]
\end{eqnarray}

This is the exact form of the bosonic partition function for 4
noninteracting particles in an arbitrary external potential. The
probability of randomly choosing a particular partition
$\phi^{i}\in\Phi$ as a function of temperature is given by

\begin{eqnarray}
\mu\left(\phi_{i}\in\Phi\right)&=&\frac{1}{Z_{B}}\textrm{dim}\left(\phi^{i}\right)\prod_{j=1}^{|\phi_{i}|}Z_{1}[\lambda_{i}^{j}\beta]^{k_{i,j}}
\end{eqnarray}

An analysis of this structure can show us the relative weight of
partition structures at various temperatures. A graph of the
contributions of the partitions of 4 to the partition function for
noninteracting particles in a 1-d harmonic trap is shown in Fig 2.

\begin{figure}
  \includegraphics[width=8cm]{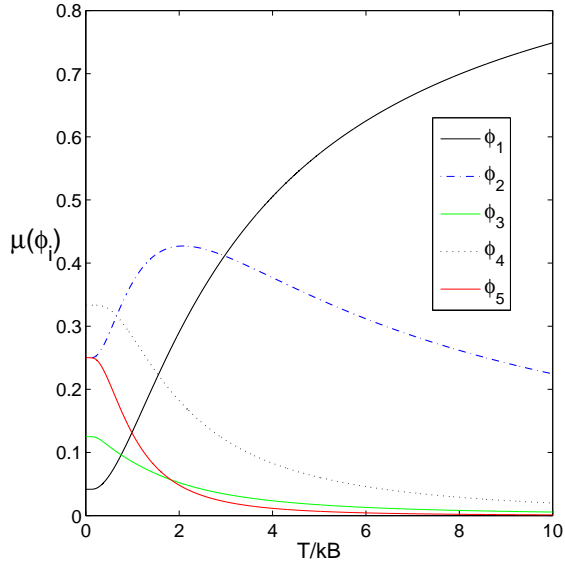}\\
  \caption{(Color online). Loop structure of the 4 particle partition function in a 1-d harmonic potential. The x-axis is
  the temperature and the y-axis the weight of the i$^{th}$ conjugacy class.\label{partition_structure.fig}}
\end{figure}

In this figure, the contributions from various elements of the
permutation group are clearly evident, showing that at high
temperatures the identity dominates, leading to the usual classical
statistics\cite{Feyn-Stat-Mech,Ceperley-1995}. Also in the
$T\rightarrow{0}$ limit, the partition function reduces to

\begin{eqnarray}
Z_{B}[\beta\rightarrow\infty]=\frac{1}{N!}\sum_{i=1}^{P(n)}\textrm{dim}\left(\phi^{i}\right)
\end{eqnarray}

which is exact for all noninteracting bosonic systems. For systems
in which the particles interact, the evaluation of partition
function elements will generally not be decomposable as a product of
single particle functions\cite{Holzmann-1999}, but the loop
structure based upon the conjugacy classes of $S_{n}$ will still be
present. However the range of problems for which Eq.(8) is exactly
solvable for interacting systems is severely limited.

\section{PIMC and Permutational Sampling}

The PIMC technique at present is the only method which renders
Eq.(8) amenable to numerical solution. However the method by which the
permutational structure is accounted for either is based upon
including only a subset of cyclic exchanges, usually up to $C_{4}$
or $C_{5}$, or the approximation of the exact permutational
structure with that of a noninteracting system. The first method
works very well in predicting properties such as the superfluid
transition temperature of helium\cite{Ceperley-1995,Ceperley-1989},
but is equivalent to the treatment of parastatistics in bosonic
systems first outlined by Green \cite{Green-Parastatistics}
\cite{Polychronakos-1996}, and does not include the full
permutational structure of the system. The second method works only
for weakly interacting systems around the superfluid transition
point as implemented in \cite{Holzmann-1999} via knowledge of the
non-interacting partition function counterpart and is unable to give
describe systems where the interactions are strong. The strong
interaction regime is important especially for the case of repulsive
interactions. It is specifically this regime in which the PIMC
method is superior as perturbation theory becomes divergent.

In recent years mathematical methods have been developed regarding
the random growth of partitions of the symmetric group, the
probability measure known as the Plancherel measure and random
matrices \cite{Borodin-2000,Polychronakos-1996,Okounkov-2000}. One
would like to ask if this abstract mathematical method could help in
the very physical application of predicting the permutational
structure of BEC's and superfluids at arbitrary temperatures. If we
denote $\Phi$ as the set of all Young tableaux with n boxes, the
probability of randomly choosing a particular partition
$\phi^{i}\in\Phi$ as a function of temperature is given by

\begin{eqnarray}
\mu\left(\phi_{i}\in\Phi\right)&=&\frac{1}{Z_{B}}\textrm{dim}\left(\phi^{i}\right)Z[\phi_{i};\beta]
\end{eqnarray}

with $Z_{B}$ equal to the total interacting bosonic partition
function. This is presumably the probability used in
\cite{Holzmann-1999} to construct partitions, which was then applied
to the case of a weakly interacting bose gas, however the way they
did this was not explicitly stated.

In the simplest implementation of a stochastic sampling method, one
would randomly select a partition and select this with the
probability given above. However, as derived by Ramanajan and Hardy,
for large n the number of partitions is given by \cite{Hardy-1918}

\begin{eqnarray}
P\left(n\right)&\simeq&\frac{1}{4\sqrt{3}n}e^{\pi\sqrt{\frac{2n}{3}}}.
\end{eqnarray}

which becomes prohibitively ineffiecient for large $n$, as the rejection rate
for transitions between uncorrelated partitions would be large.

We propose a set of operators which enable us to construct a Markov
chain through the conjugacy classes of $S_{N}$. A similar formalism
can be found in Borodin and Olshanski
\cite{Borodin-2004,Borodin-Markov} in constructing partitions,
identifying an operation which maps between partitions. Again
consider a partition as the weakly decreasing sequence

\begin{eqnarray}
\bar{\phi^{i}}&=&\{\bar{\lambda}^{i}_{1},\bar{\lambda}^{i}_{2},...,\bar{\lambda}^{i}_{j},...,\bar{\lambda}^{i}_{\rho-1},\bar\lambda^{i}_{\rho}\}
\end{eqnarray}

but with the restriction that at least one element, namely
$\bar{\lambda}^{i}_{\rho}$, has the value of 1. Consider two
operators $a_{j}$ and $\bar{a}_{j}$, such that

\begin{eqnarray}
\nonumber a_{j}\phi^{i}&=&\{\lambda^{i}_{1},...,\left(\lambda^{i}_{j}+1\right),...,\lambda^{i}_{\rho-1}\}\\
\nonumber &=&\phi^{k}\\
\nonumber\bar{a}_{j}\phi^{i}&=&\{\lambda^{i}_{1},...,\left(\lambda^{i}_{j}-1\right),...,\lambda^{i}_{\rho},\lambda^{i}_{\rho+1}\}\\
 &=&\phi^{m}
\end{eqnarray}

such that

\begin{eqnarray}
\nonumber|a_{j}\bar{\phi}_{i}|&=&|\bar{\phi}_{i}|-1\\
|\bar{a}_{j}\bar{\phi}_{i}|&=&|\bar{\phi}_{i}|+1.
\end{eqnarray}

These operators create and destroy $C_{1}$ subgroups of the
conjugacy classes, in the process creating a new partition which is
also a conjugacy class of the relevant permutation group. The action
of these operators is to effectively move up and down a graph with
$P(n)$ vertices of the partitions of $S_{N}$, leaving the sum of the
elements of partitions invariant (cf. fig.3).

\begin{figure}
  \includegraphics[width=3cm]{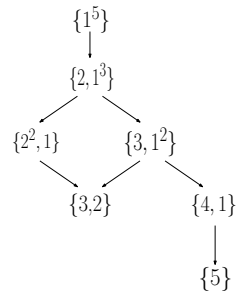}\\
  \caption{Neighboring partitions of $S_{5}$ under the action of $a$ and $\bar{a}$}\label{part.s5}
\end{figure}

If we denote partitions which are related via a single application
of $a_{j}$ or $\bar{a}_j$ as neighouring partitions, we can
construct a probability measure over neighbouring partitions to give
a formula for the transition probability under $a_{j}$ and
$\bar{a_{j}}$ as

\begin{eqnarray}
\nonumber p^{\downarrow}\left(\phi_{i};a_{j}\phi_{i}\right)&=&\frac{\textrm{dim}\left(a_{j}\phi_{i}\right)}{\sum_{j}\textrm{dim}\left(a_{j}\phi_{i}\right)}\\
p^{\uparrow}\left(\phi_{i};\bar{a}_{j}\phi_{i}\right)&=&\frac{\textrm{dim}\left(\bar{a}_{j}\phi_{i}\right)}{\sum_{j}\textrm{dim}\left(\bar{a}_{j}\phi_{i}\right)}
\end{eqnarray}

One can easily see that these transition probabilities satisfy the
criterion that the sum of probabilities over neighbours is equal to
one. In the construction of a Markov chain through permutation
space, one could suppose that these are the correct transition
probabilities to be used. Ergodicity, in the sense that the correct
weight of each partition will be reproduced,  is not assured via
these relations. It is not however difficult to prove that all
partitions are accessible under this scheme. If we consider the
partition denoting the identity of $S_{N}$ as $\{1\}_{N}$, then we
may construct any other partition using a repeated application of
$a_{j}$, that is

\begin{eqnarray}
\{\lambda_{1},\lambda_{2},...,\lambda_{\rho}\}&=&\prod_{i=1}^{\rho}a_{i}^{\lambda_{i}}\{1\}_{N}.
\end{eqnarray}

Since we are able to perform the reverse of this operation, that is
reach the identity from any partition, via the use of $\bar{a}_{j}$,
all partitions are connected. Once a partition is chosen then it can
be accepted or rejected via the Metropolis scheme\cite{Metropolis}.
One hinderance to the successful application of these transition
probabilities is that there does not exist any formula for the sum
over neighbours which would be relatively easy to implement on a
computer within a Monte Carlo code, especially for systems
containing a large number of particles.

An alternative option which is relatively easy to implement would be
to give all neighbours the same weight. Then one may choose a
particular element of a partition to act upon with the probabilities

\begin{eqnarray}
p^{\downarrow}\left(\phi_{i};a_{j}\phi_{i}\right)&=&\frac{k_{j}}{|\bar{\phi}_{i}|-1}\\
p^{\uparrow}\left(\phi_{i};\bar{a}_{j}\phi_{i}\right)&=&\frac{k_{j}}{|\bar{\phi_{i}}|-k_{i,\gamma}}.
\end{eqnarray}

The action of $\bar{a}_{j}$ on a partition element is to give
$\bar{a}(\lambda)\rightarrow(\lambda-1,1)$ and as such the
normalisation is the number of elements in a partition which are not
equal to one. The action of $a_{j}$ on a partition element is to
give $a(\lambda,1)\rightarrow(\lambda+1)$ such that at least one
element must have value one, and as such the normalisation is the
number of elements in a partition less one.

It was essentially this use of $a_{j}$ exclusively by Boninsegni in
\cite{Boninsegni-2005} to construct permutation cycles for the
superfluid phase of $^{4}$He, containing 64 atoms. However in this
work permutation cycles were constructed at each sampling move from
the identity. Again, from Eq.(28) we can see that each permutation
cycle will be accessible via this scheme, however it is not
guaranteed that the correct weights will be reproduced.

\section{Computation and Discussion of Results}
As a test of the proposed scheme with the analytic results given in
Fig. 2, we performed a path integral Monte Carlo calculation on 4
bosons in a 1-d harmonic trap. We used essentially the method as
outlined in\cite{Ceperley-1995}, in the primitive
approximation\cite{Trotter_Number}.

The Trotter number was chosen such that the single particle energy
was comparable to the analytic expression
$\langle{E}\rangle=\frac{1}{2}\coth\left(\frac{1}{2}\beta\right)$,
however small enough that chosen permutations were accepted.

We implemented two schemes for permutation sampling. In the first,
the results of which are summarised in Fig. (4), particle labels
were randomly shuffled at each move and were accepted or rejected
according to the Metropolis scheme. This was used primarily as a
test for the validation of the code. In the second scheme we
constructed a random walk through permutation space based upon the
transition probabilities given in Eqs. (29-30) between partitions.
At each move a partition list is created and an up or down move
($a_{j}$ or $\bar{a}_{j}$) proposed with equal probability. In the
case that a partition is at the end of a path (cf. Fig. (3)), then
the algorithm forces a move to the next connected vertex. In the
case of $a_{j}$, once a partition element is chosen, then the
nearest single loop structure is found such that the acceptance
probability in the following Metropolis sampling move is maximised.
The results of this method are shown in Fig. 5.

As can be seen, randomly shuffling particle labels reproduces the
analytic decomposition of the partition function very well. We would
however expect this method to become diminishingly inefficient as
the particle number increased, and as the confining potential
becomes significantly weaker, where the thermal wavelength of the
particles will no longer be comparable to the mean particle
separation, and thus would not be a viable option for a general PIMC
code. Further the results of our method show that even though every
partition is accessible via this algorithm, this does not guarantee
that the method will reproduce the analytical result of Eq.(20).

\begin{figure}
  \includegraphics[width=9cm]{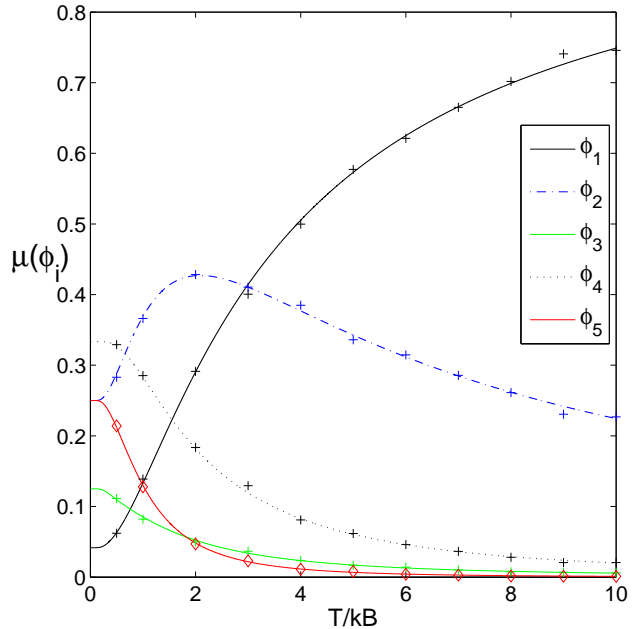}\\
  \caption{(Color online). Partition structure of four bosons in a 1-d harmonic trap from PIMC
    calculations, with partitions created via a random shuffling of particle labels. The solid line
    is the exact answer as given by Eq.(20)}\label{part.s5}
\end{figure}

\begin{figure}
  \includegraphics[width=9cm]{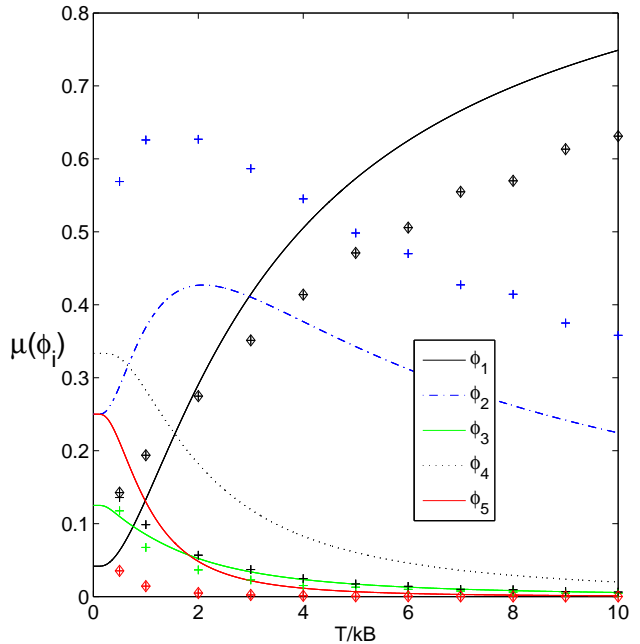}\\
  \caption{(Color online). Partition structure of four bosons in a 1-d harmonic trap
    with partitions created the new method. The solid line
    is the exact answer as given by Eq.(20)}\label{part.s5}
\end{figure}

An interesting feature of the results of the two methods of
permutational sampling investigated is that, whilst they did not
coincide with regard to the predicted permutation cycle structure,
the resultant energy of the two methods was nearly identical (cf.
Fig. (6)).

\begin{figure}
  \includegraphics[width=9cm]{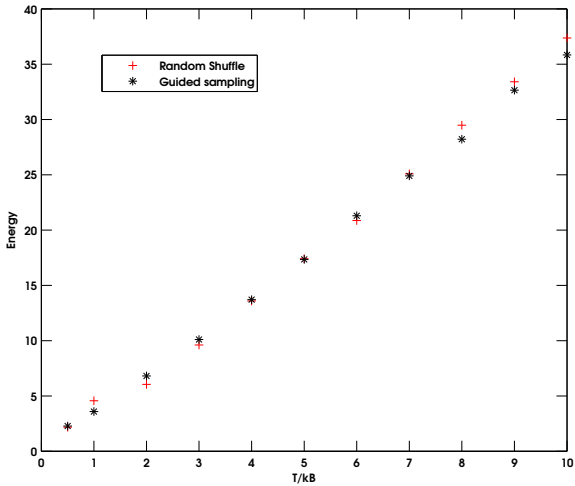}\\
  \caption{(Color online). Energy as a function of temperature for 4-particles bosonic particles
    using two different methods for sampling permutation space. The statistical errors in these
    calculations are beyond the resolution of this graph.}
\end{figure}

Although the new method proposed was unable to reproduce the exact
analytic form of Eq.(20) for this relatively basic model, we would
expect that as the number of particles is increased that this would
become more accurate. As the number of paths between vertices of the
partition graph is increased, a walk on this graph will not be
required to pass through a disproportionate number of low weight
vertices. Our results show that the weight of the larger cycle
lengths is underestimated, which is reflected by the fact that they
lie on the end points of the graph of $S_{4}$. The larger cycle
lengths contribute to quantities such as the mean winding number
which derived from the mean squared winding
number\cite{Ceperley-1989}. This method does have the advantage over
other methods in that the \textit{only} approximation is the form of
the transition probabilities. After a period of equilibration, where
the cycle structure is constructed from the identity, we will be
able to sample the local permutational structure by using $a_{j}$
and $\bar{a}_{j}$ without any \textit{a priori} knowledge of cyclic
structure of the partition function.

As mentioned earlier, the number of conjugacy classes of $S_{n}$ has
the asymptotic form

\begin{eqnarray}
P\left(n\right)&\simeq&\frac{1}{4\sqrt{3}n}e^{\pi\sqrt{\frac{2n}{3}}}.
\end{eqnarray}

which would provide the associated computational sampling frequency as a
function of particle number. However the overhead required for the
construction of partitions is minimal, with only knowledge of the
current permutational structure and the proposed structure required,
and thus memory requirements kept to a minimum. Further it is not
necessary that the entire graph of partitions be sampled as many
configurations will have only minor contributions, these depending
on the temperature. After an initial period of equilibration, the
process will sample the local neighbourhood of the highest weight
permutations, those which will contribute the greatest to
observables. Although the intrinsic $n!$ scaling involved in the
sampling of permutation space has largely been
overcome, they have been replaced by algorithms scaling exponentially
\cite{Ceperley-1995,Boninsegni-2005,Holzmann-1999}, which is still
problematic for large $n$. A recent proposal for a size independent
algorithm has been reported by Boninsegni \textit{et. al.} in
\cite{Boninsegni-PRL} which also looks promising for investigations
into large systems and calculating off-diagonal elements of the
density matrix.

\section{Conclusion}

To summarize, we suggest the use of one loop operators acting upon
partitions for the construction of a Markov chain through
permutation space. We have derived a form of the partition function
for bosonic systems in the canonical ensemble based upon the
irreducible representations of the symmetric group of n objects. It
was shown that the partition function can be decomposed into a sum
over the conjugacy classes of $S_{n}$ which defines a probability
measure over Young tableaux which is a function of temperature.

There have been various proposals for a more transparent method of
sampling permutation space in the recent literature both for bosonic
and fermionic systems
\cite{Boninsegni-2005,Lyubartsev-93,Holzmann-1999,Boninsegni-PRE,Boninsegni-PRL},
which appear to be successful in predicting properties of interacting quantum
systems. However there still does not exist an unambiguous method
for the sampling of permutation space at the same level of rigor as
for calculating expectation values of continuous probability
measures. One may expect that a multilevel Metropolis scheme could
be developed whereby the partition structure at finite temperature
is reproduced exactly, giving a more accurate sampling of the
partition function. Any scheme that purports to full sampling of the
permutational structure of atomic gases within path integral Monte
Carlo should be able to reproduce the analytic form for the non
interacting case. A more rigorous formulation for the construction
of discrete probability measures amenable to Monte Carlo methods may
lead to some very interesting new observations.

An interesting mathematical question also arises which we wish not
to explicitly address here, but may be a useful connection between
the theory of indistinguishable many-body systems and a large body
of work in probability theory and the random growth of Young
tableaux. Is one able to construct a Markov chain via the use of
$a_{j}$ and $\bar{a}_{j}$ in the canonical ensemble that in the
infinite time limit will give probabilities distributed according to
equation 20? Further, what is the limit shape of Young tableaux in
the $N\rightarrow\infty$ limit as a function of
temperature\cite{Borodin-2000,Okounkov-2000}? As
$T\rightarrow\infty$ we expect this to be
$x\theta\left(x+\frac{1}{2}\right)$, where $\theta$ is the Heaviside
step function, as the identity will become dominate in this limit.
In the $T\rightarrow{0}$ limit, where the partition function reduces
to Eq.(21), the limit shape in the grand canonical ensemble has been
established and as such one would expect a continuous transformation
between these limit shapes, the mapping being dependant on
temperature. This will be related to finding eigenvalues of random
matrices as a function of temperature. We hope to investigate this
further in a future paper.

\begin{acknowledgments}
R. Springall would like to thank all attendees of the ICE-EM
graduate school in mathematical physics held at the University of
Queensland in July of 2006. M. Per acknowledges the support of the
Australian Research Council.
\end{acknowledgments}

\bibliography{bosonic_path_integral}

\begin{thebibliography}{30}
\bibitem[{Tro()}]{Trotter_Number}
\emph{\bibinfo{title}{This is $e^{A+B}=e^{A}e^{B}$. there are more accurate
  forms available based upon the proper expansion in $\tau$ however we wish
  only here to look at the permutation structure which should be relatively
  independent of this}}.
\end{thebibliography}

\end{document}